\documentclass[final,authoryear,5p]{elsarticle}

\usepackage{epsfig}

\usepackage{amssymb}

\usepackage[ps2pdf,%
a4paper=true,%
breaklinks=true,%
colorlinks=true,%
pdfauthor={First Author et al.},%
pdftitle={Template for manuscripts in Advances in Space Research}%
]{hyperref}

\journal{Advances in Space Research}

\begin{document}

%%%%%%%%%%%%%%%%%%%%%%%%%%%%%%%%%%%%%%%%%%%%%%%%%%%%%%%%%%%%%%%%%%%%%%%%%%%%%
%% Frontmatter
\begin{frontmatter}


% Use the tnoteref command within \title and fnref within \author or \address for footnotes;
% use the corref command within \author for corresponding author footnotes;
% \author{Name\corref{cor1}\fnref{label2}}
% \address{Address\fnref{label3}}

\title{Voyager observations in the distant heliosheath: An analogy with ISEE-3 observations in the deep geomagnetic tail}

% Use optional labels to link authors explicitly to addresses:

\author{Ian G. Richardson\corref{cor}\fnref{footnote2}}
\address{CRESST and Department of Astronomy, University of Maryland, College Park, 20742, USA}
\cortext[cor]{Corresponding author}
\fntext[footnote2]{Also Code 661, NASA Goddard Space Flight Center, Greenbelt, MD 20771, USA}
\ead{ian.g.richardson@nasa.gov}

% Url can be given like this:
% \ead[url]{http://www.elsevier.com/wps/find/authorsview.authors/latex}

%\author{Second Author and Third Author\fnref{footnote3}}
%\address{Address of the second and third authors}
%\fntext[footnote3]{Additional information about the second and third authors}
%\ead{more@email.addresses}

%\author{More Authors\fnref{footnote4}}
%\address{Address of the co-authors}
%\fntext[footnote4]{Additional information about the co-authors}
%\ead{more@email.addresses}

\begin{abstract}
We suggest an analogy between energetic particle and magnetic field observations made by the Voyager 1 spacecraft in the distant heliosheath at 122~AU in August 2012, and those made in the distant geomagnetic tail by the ISEE~3 spacecraft in 1982--1983, despite large differences in the time and distance scales.  The analogy suggests that in August, 2012, Voyager 1 may not have moved from the anomalous cosmic ray (ACR)-dominated heliosheath into the interstellar medium but into a region equivalent to the ``lobes" of the geomagnetic tail, composed of heliospheric field lines which have reconnected with the interstellar medium beyond the spacecraft and so are open to the entry of cosmic rays, while heliospheric particles (e.g., ACRs) are free to escape, and which maintain a $\sim$Parker spiral configuration.  The heliopause, analogous to the magnetopause forming the outer boundary of the lobes, may then lie beyond this so-called ``heliocliff".  Even if this analogy is incorrect, the remarkable similarities between the energetic particle and magnetic field observations in these very different regions are worth noting.

\end{abstract}


\end{frontmatter}

\parindent=0.5 cm

%%%%%%%%%%%%%%%%%%%%%%%%%%%%%%%%%%%%%%%%%%%%%%%%%%%%%%%%%%%%%%%%%%%%%%%%%%%%%
%% Main text

\section{Introduction -- Voyager Observations}
\begin{figure}
% \center\includegraphics[width=22pc]{C:/figures/Owen-voyager.eps}
\center\includegraphics[width=22pc]{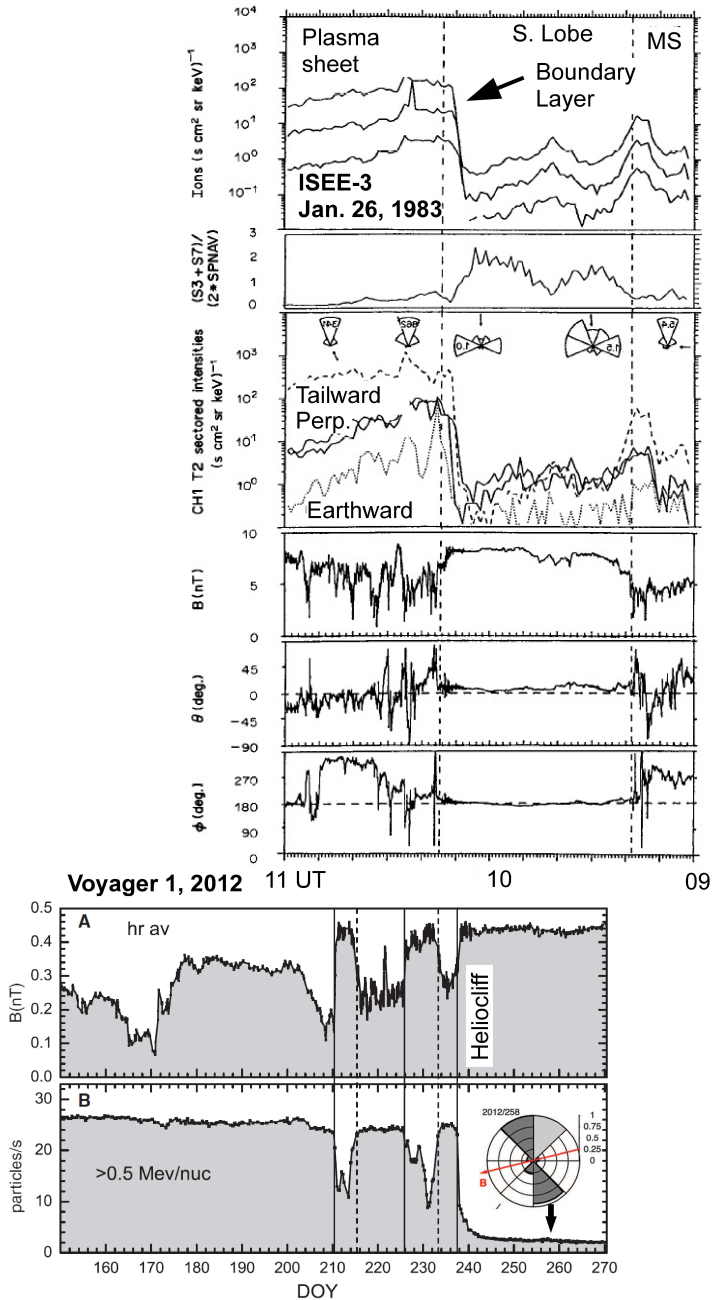}

 \caption{ISEE--3 EPAS energetic ion (35-56, 56-91, 91-147 keV) and magnetic field observations (top part; adapted from Owen et al. [1991b]) for 09-11~UT on January 26, 1983 at 213~$R_e$ downtail, but with the time axis reversed in order to illustrate the similarity of particle intensity and magnetic field variations at the plasma sheet-lobe transition and those observed by V1 at the heliocliff (bottom part; Burlaga et al. [2013]). Sectored particle distributions peaked perpendicular to the magnetic field direction are seen both in the low intensity particles in the lobe and $\sim20$~days following the heliocliff at V1 (Krimigis et al., 2013).   A crossing of the magnetopause from the lobe to the magnetosheath is also shown at ISEE-3.  Additional details are given in the text.   
}
 \label{i3v}
 \end{figure}

Observations made by the Voyager 1 (V1) spacecraft in the distant heliosheath at 122~AU in August 2012, illustrated in the bottom part of Figure~\ref{i3v} \citep{bur13}, showed unanticipated dramatic, abrupt, temporary decreases in the intensity of ``anomalous" cosmic rays (ACRs; bottom panel of the V1 observations) after several years of near-constant intensities as the spacecraft crossed the heliosheath, followed by a precipitous and persistent decrease \citep{kri13}.  The temporary decreases in ACR intensity were accompanied by temporary increases in the intensity of galactic cosmic rays (GCRs; not illustrated here), with a persistent GCR enhancement  commencing at the final ACR decrease \citep{st13}.  There were also temporary increases in the magnetic field strength at the times of the ACR decreases/GCR increases (upper panel of the V1 observations in Figure~\ref{i3v}), with a final and persistent increase of the field strength commencing at the final ACR decrease (right-hand vertical solid line).  \citet{bur13} reported that the magnetic field direction showed little change at the time of the ACR and GCR variations, remaining closely aligned with the $\pm$T (transverse) direction expected for the heliospheric magnetic field.  This observation suggests that the final, dramatic change in the particles and magnetic fields -- the ``heliocliff" -- was not the much-anticipated crossing of the heliopause into the interstellar medium (ISM), where a different field orientation would be expected.  On the other hand,  V1 plasma wave observations \citep{gur13} indicate relatively high plasma densities outside the heliocliff (V1 does not have an operating plasma instrument) that are consistent with those expected in the ISM, leading the Voyager team to conclude that V1 did in fact cross the heliopause in August, 2012.  This view has been challenged by  \citet{fg14} and \citet{gf15}, who suggest that these high densities are instead associated with compressed solar wind and that the heliopause has yet to be reached.

The intent of this paper is simply to point out some interesting similarities between the V1 observations in August 2012 and energetic ion and magnetic field observations made during the exploration of the distant ($\le 200 R_e$) geomagnetic tail by the ISEE~3 spacecraft in 1982--1983 \citep{tvr84}, albeit at vastly different temporal (days \textit{vs.} minutes) and spatial (AU \textit{vs.} $R_e$) scales.  The similarities suggest that V1 may still be in the heliosphere following the heliocliff.

\section{ISEE~3 Observations}
\begin{figure}
% \center\includegraphics[width=25pc]{C:/figures/Hughes-ms-2.eps}
\center\includegraphics[width=20pc]{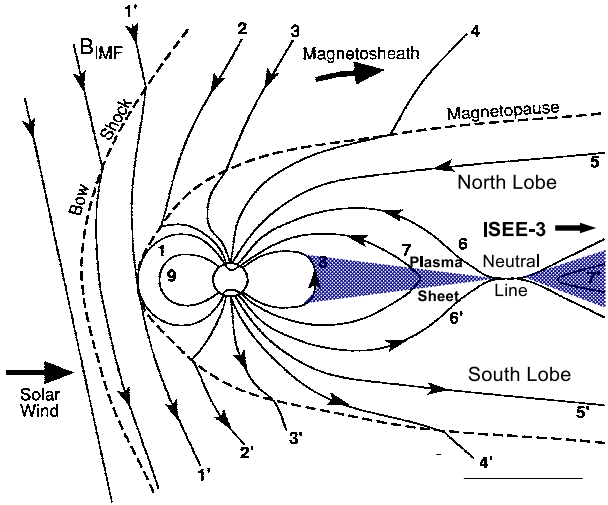}

 \caption{Schematic of the magnetosphere for southward IMF (adapted from Hughes, 1995) showing field lines (numbered 1' to 4 and 4' in sequence) dragged into the tail following dayside reconnection; 5, and 5' are lobe field lines, connected to the Earth and open to the solar wind.  Hatched shading indicates the plasma sheet in the tail center plane resulting from particle acceleration due to reconnection at the tail neutral line.  As indicated, ISEE-3 was typically tailward of this neutral line when in the distant tail.}
 \label{ms}
 \end{figure}

The top part of Figure~\ref{i3v} shows a 2-hour interval of ISEE-3 energetic particle observations, from the Energetic Particle Anisotropy Spectrometer (EPAS) (\citet{b78}; \citet{o91b}), and magnetic field data, made when the spacecraft was 213~$R_e$ downtail from Earth.  This is based on Figure~6 of \cite{o91b} but that figure has been flipped from right to left (i.e., the time axis is reversed) to compare better with the V1 observations in the bottom part of Figure~\ref{i3v}.  To place the ISEE-3 observations in context, Figure~\ref{ms} shows a simple schematic of the magnetotail (adapted from \citet{h95}) that is sufficient for the purposes of this paper.  The tail consists first of two ``lobes" composed of magnetic field lines (numbered 5 and 5') that are connected to the Earth at one end and are open to the solar wind at the other.  These field lines have previously reconnected with the interplanetary magnetic field (IMF)  at the dayside (field line 1'; a southward IMF, favorable for reconnection, is shown) before being dragged over the polar caps into the tail (2, 3, 4, 2', 3', 4') by the solar wind flow past the Earth \citep{d61}, which also aligns the lobe field lines along the tail. The north and south tail lobes are separated by a plasma sheet (hatched region in Figure~\ref{ms}) formed by the reconnection of lobe field lines at a neutral line.  Field lines 6 and 6' map to the neutral line, as do the equivalent field lines downtail from the neutral line.  Plasma accelerated at the neutral line flows away from the neutral line both toward the Earth and down the tail.  In the deep tail, ISEE--3 was usually tailward of the neutral line and observed predominantly tailward flows in the plasma sheet \citep{z84}.  In the plasma sheet, thermal plasma pressure is dominant, and the magnetic field is weak and variable in direction.  Finally, the magnetopause (MP) marks the boundary between the lobes and the ambient solar wind which has passed through the bow shock and is not connected to the magnetosphere.  Dynamical processes in the tail also occur, such as the formation of plasmoids during substorms which were detected by ISEE-3 in the deep tail (e.g., \cite{rc85}, \cite{mh92}).  See \cite{san12} for a recent review of substorm dynamics.

When in the deep tail, ISEE-3 sampled different structures as the tail moved across the spacecraft due to tail dynamics and flapping of the tail in response to changes in the solar wind velocity \citep{h86}.  During the (very typical) two-hour period illustrated in the top part of Figure~\ref{i3v}, ISEE-3 was in the plasma sheet to the left of the first dashed vertical line, characterized by a magnetic field that is variable in intensity and direction (the bottom three panels in the top part of the figure show the magnetic field strength and polar and azimuthal angles in GSE coordinates), and by a population of energetic ions observed by EPAS; direction-averaged intensities in the three lowest EPAS energy channels (35-56, 59-91 and 91-147~keV total energy) are shown in the top panel.  The third panel shows 35-56~keV ion intensities from EPAS telescope 2 (inclined at $30^o$ from the ecliptic) in various flow directions (see \citet{o91b} for further details). Also illustrated are sample pie plots of particle counts accumulated in 64 s intervals in the 8 azimuthal sectors of this telescope as the spacecraft rotated about a spin axis oriented perpendicular to the ecliptic.  As discussed by \cite{r87}, ion anisotropies observed by ISEE-3 in the distant tail plasma sheet were generally consistent with Compton-Getting anisotropies arising from convection with the thermal plasma.  In the case shown here, the flow is tailward as can be seen in the pie plots, where the dominant sector facing the top of the figure corresponds to ions flowing tailward, and also in the third panel, where the tailward--flowing ion intensity is dominant. 

In the centeral period of the ISEE-3 observations in Figure~\ref{i3v} (between the vertical dashed lines), ISEE-3 was in the south lobe, characterized by strong, low variance, tail-aligned fields ($\phi \sim180^o$, $\theta \sim0^o$) and energetic ion intensities around two orders of magnitude lower than in the plasma sheet.  The pie plots in this region show that these ions have ``pancake" angular distributions peaked at $\sim90^o$ to the magnetic field direction. Such angular distributions are a pervasive feature of tail lobe ions as discussed and modeled by \citet{o91a} and \citet{o91b}.  This is also illustrated in the second panel of the ISEE-3 observations in Figure~\ref{i3v} which shows the ratio of the sum of the 35-56~keV ion counts in the two sectors of telescope 2 viewing perpendicular to the lobe field to twice the spin-averaged sector count.  This ratio is enhanced above 1 throughout most of the lobe interval, indicating that above--average count rates occurred in the perpendicular sectors.  Also, in the third panel, intensities perpendicular to the field are generally dominant in the lobe. Pancake distributions are present in the lobes because ions with velocity components along the tail are lost along the open lobe field lines (after reflection near the Earth if initially moving Earthward), leaving a small remnant population of $\sim90^o$ pitch angle ions which is relatively stable in the low variance lobe fields. %Solar energetic particles also can readily enter the tail lobes (\cite{o91a}, \citet{o91b}).  

In the right-hand period of the ISEE-3 observations in Figure~\ref{i3v}, ISEE-3 has crossed the MP into the magnetosheath.  The magnetic field is again more variable since this is solar wind plasma that has passed through the bow shock.  Energetic ions show a tailward flow due to convection with the solar wind flow, together with, in this case, a modest intensity enhancement in the vicinity of the MP.  We should emphasize that the observations in Figure~\ref{i3v} are very typical of the different deep tail regions sampled by ISEE-3 predomantly due to transverse motions of the tail relative to the spacecraft, and the fact that the time axis has been reversed here does not impact their interpretation.    

\section{Comparing V1 and ISEE-3 Observations}
Comparing the V1 and ISEE-3 observations in Figure~\ref{i3v}, and notwithstanding the factor of 1440 difference in the time intervals shown (2~hours vs. 120 days), we point out several close resemblances: (1) The drop in the energetic ion intensity associated with the transition from the plasma sheet into the lobe, and that at the heliocliff; (2) The onset of stronger, low variance magnetic fields on moving from the plasma-dominated plasma sheet into the lobe, similar to the transition observed as V1 crossed the heliocliff from the ACR-dominated region; (3) The presence of low intensities of ions with pancake ion distributions peaked perpendicular to the magnetic field in the lobe, and the similar pancake distributions found in the declining ACR intensity at V1 following the heliocliff.  These distributions persisted for some time, in particular at higher energies \citep{kri13} -- the V1 sectored ion intensity plot shown in the bottom part of Figure~\ref{i3v} is from 20 days following the heliocliff; no measurements were made in the light-shaded sector, and the arrow indicates the magnetic field direction (which is $\sim$orthogonal to the tail lobe field in the figure);  (4) The lobe field direction is determined by the solar wind flow past the Earth, while the transverse Parker spiral field at V1 is  caused by the rotation of the heliosphere, and (5) Solar energetic particles can freely enter the open lobes of the geomagnetic tail, as modeled by \cite{o91b}, and analogously, after the heliocliff, V1 moved into a region into which GCRs can freely enter.  

The similarity in these observations suggests that on crossing the heliocliff, V1 moved into a region analogous to the tail lobes, consisting of field lines that have reconnected with the interstellar field downstream of the spacecraft but are still connected to the heliosphere at the sunward end.  The presence of field lines beyond the heliocliff that are open to the ISM has been suggested in several of the papers cited above, and will be discussed further below, so this is certainly not an original suggestion.  Nevertheless, Figure~\ref{i3v} illustrates how phenomena observed in one situation may help in the interpretation of somewhat similar phenomena found in very different circumstances.  Comparison of the observations suggests furthermore that the heliopause (analogous to the magnetopause) may lie beyond the heliocliff.  

There are also several differences in the Voyager and ISEE-3 observations beyond the time and distance scales.  The strongly tailward streaming ions observed by ISEE-3 in the plasma sheet in the top part of Figure~\ref{i3v} are due to convection with high-speed plasma, often reaching several hundred km/s \citep{r87} compared with near-zero solar wind speeds in the heliosheath that are inferred from the nearly isotropic ion distributions at V1 \citep{d12}.  ISEE-3 also typically observed an energetic ion ``plasma sheet boundary layer" (PSBL), evident in Figure~\ref{i3v} extending a short distance beyond the edge of the plasma sheet onto lobe-like field lines (i.e., extending to the right of the left-hand vertical dashed line) .  An associated small diamagnetic depression in the magnetic field, and waves driven by the right-hand resonant ion beam instability \citep{t85}, are also typically present and may just be discerned in the figure.  Tailward ion anisotropies in the PSBL are typically larger than expected from convection with the local plasma, indicating that these ions are streaming freely through the thermal plasma \citep{r87}.   As discussed by \cite{rc85}, the PSBL in the deep tail is formed by energetic particles accelerated at the neutral line that stream away from the neutral line faster than the thermal plasma sheet plasma and populate a wedge of field lines overlying the plasma sheet that maps back to the NL.  The thickness of this wedge is determined by particle speed, with faster particles producing a thicker boundary layer; the last field line that has reconnected at the neutral line (i.e., the ``separatrix"; field lines 6, 6', and the corresponding field lines downtail of the NL in Figure~\ref{ms}) marks the topological boundary of the PSBL for infinite speed particles.  %Hence there is energy dispersion at the edge of the PSBL \citep{rc85} (not evident at the scale of Figure~\ref{i3v}).  
It should be emphasized that the PSBL is not simply a gyroradius scale effect due to plasma sheet particles gyrating into the lobe but is a consequence of field line connection to, and particle acceleration at, a neutral line upstream of the spacecraft.  The intensity decrease at the edge of the PSBL is then on a gyroradius scale, and $B\times\bigtriangledown N$ anisotropies may be observed, transitioning to the pervasive, low intensity, lobe ions with pancake distributions \citep{rc85}.  Another obvious difference is that the tail is symmetric about the plasma sheet, whereas the Voyager analogy corresponds to ``half" of the tail, with the plasma sheet playing the role of the heliosheath. 

  \begin{figure}
 %\center\includegraphics[width=30pc]{C:/figures/richcowley-2.eps}
 \center\includegraphics[width=20pc]{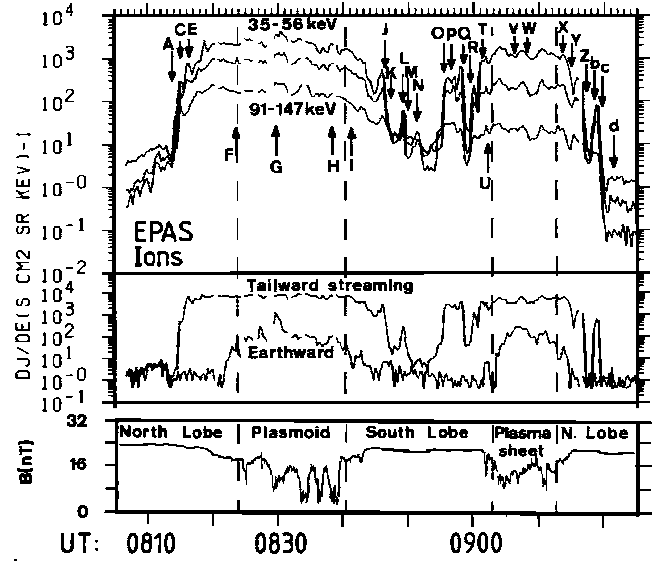}

 \caption{Observations (adapted from Richardson and Cowley, 1985) for an interval on March 25, 1983 when ISEE-3 was 109~$R_e$ downtail and passed from the north lobe through the plasma sheet (during passage of a plasmoid) to the south lobe and returned via the plasma sheet to the north lobe.  Enhancements of tailward streaming ions were observed in the plasma sheet and in the adjacent plasma sheet boundary layer on lobe-like field lines.  Brief oscillations in the ion intensity are due to the edge of the region enhanced intensity moving back and forth across the spacecraft, as discussed by Richardson and Cowley (1985); letters refer to ion angular distributions in Figure~10 of that paper.}
 \label{rc}
 \end{figure}

ISEE-3 observations also suggest a possible interpretation of the anti-correlated variations in ion intensity and magnetic field strength preceding the heliocliff.  Figure~\ref{rc}, adapted from \cite{rc85}, shows an interval in which ISEE-3 (at 109~$R_e$ downtail) moved from the north lobe through the plasma sheet (which in this case included passage of a plasmoid; see \cite{rc85} for details) into the south lobe and then returned to the plasma sheet and north lobe.  The top panel shows EPAS ion intensities in 3 energy channels (35--56, 56--91, 91--147~keV), while the other panels show the intensities of tailward- and earthward-streaming 35--56~keV ions in Telescope 2 and the magnetic field intensity, which again illustrates the contrast between the weaker, variable fields in the plasma sheet and the persistent, strong fields in the lobes.  The PSBL, with strongly tailward-flowing ions, can be seen prominently on lobe-like field lines adjacent to each boundary crossing of the plasma sheet (dashed vertical lines).  Of particular interest here are the brief, temporary increases and decreases in the ion intensities that occur before the final crossing into or out of the PSBL, e.g., at times J--N, O--T, X--c indicated in the figure (see Figure~10 of \cite{rc85} for sectored ion distributions at each time).  As discussed by \cite{rc85}, these intensity fluctuations were associated with the spacecraft repeatedly briefly sensing the edge of the PSBL, probably due to tail motion, before the final crossing into or out of the PSBL.  These motions of the edge of the PSBL were confirmed by comparing particle intensities in the six EPAS sectors (two from each of the three EPAS telescopes) viewing perpendicular to the lobe magnetic field (see Figure~4 of \cite{rc85}).  Such intensity fluctuations/PSBL motions were commonly observed by ISEE-3 at the transition between plasma sheet and lobe.   

Turning to the V1 observations, the anti-correlations between the temporary decreases in ACR intensity and simultaneous increases in field intensity similarly may be due (as also suggested by \citet{qw13}) to the spacecraft at least twice temporarily approaching and crossing the heliocliff and sensing the related transitions in field and ion intensities (first two solid vertical lines in the bottom part of Figure~\ref{i3v}) before retreating again (dashed vertical lines), presumably due to motion of the heliocliff, before making the final and complete crossing (last solid vertical line).  There may also be a brief approach to the heliocliff midway between the first dashed and second solid vertical lines suggested by a brief spike in the field intensity.  The ACR and GCR observations of \cite{kri13} and \cite{st13} appear to support this picture since during at least the second possible approach to/crossing of the heliocliff, pancake ion distributions appear to be present (as discussed above, a feature of the lower intensity ions following the heliocliff), and there are also temporary increases in the GCR intensity associated with each approach that however do not reach the final and persistent GCR intensity observed following the heliocliff.  On the other hand, \citet{f15} note that the intensities observed in two sectors of the V1 LET detecting ions with gyro centers closer to and further from the Sun relative to the spacecraft declined and increased in sequence during the particle intensity/field fluctuations.  This suggests that sequence of structures passed over the spacecraft rather than an advancing then retreating front.  However, the presence of, for example, corrugations on the front (which may also, at least locally, have a normal that is inclined to the radial direction) might be a way to reconcile the anisotropy observations with a single boundary.

\section{Summary and Discussion}
We have noted similarities between V1 energetic particle and magnetic field observations near the crossing of the heliocliff in August, 2012 and at the transition between the plasma sheet and lobes in the distant geomagnetic tail observed by ISEE-3, despite the vast differences in size and time scales.  We suggest that the heliocliff is a boundary between an ACR-dominated region of the heliosheath, analogous to the thermal plasma-dominated tail plasma sheet, and a region of field lines that remain connected to the heliospheric field but have reconnected with the interstellar field beyond V1 that is analogous to the lobes of the geomagnetic tail.  The analogy then suggests that the heliopause is located not at the heliocliff but some distance beyond, analogous to the magnetopause.  The temporary variations in ACR, magnetic field and GCR intensities observed by V1 ahead of the heliocliff may be due to slight in and out motions of the heliocliff causing V1 to briefly sense the changes in these intensities as it approaches the heliocliff then retreats; similar motions were frequently observed by ISEE-3 at the edge of the plasma sheet boundary layer.  However, recently-reported V1 ion anisotropies may suggest a more complicated scenario.  Even if this analogy turns out to be invalid, it is nonetheless interesting to note the remarkable similarities between these disparate sets of observations. 

Since V1 is headed towards the nose of the heliosphere, the suggested lobe-like region following the heliocliff is clearly unrelated to the tail of the heliosphere.  It is also not clear from the single spacecraft observations whether the heliocliff is a large-scale boundary, or the crossing of a relatively localized structure, possibly dynamic in nature rather than static.

In the geomagnetic tail, we have discussed how an energetic particle plasma sheet boundary layer layer is formed that maps to a neutral line Earthward of the spacecraft and that the boundary of this layer is a topological boundary formed by the field separatrix mapping to the neutral line.  An analogy on a large scale might be with the blunt termination shock scenario of  \cite{ms12} in which the heliocliff is associated with the crossing of the last field line (also a topological boundary) that connects to the shock and is filled with ACRs. \cite{sm13} have suggested that reconnection between heliospheric and interstellar fields analogous to a flux transfer event (and consistent with a lobe-like topology) might account for the V1 observations.  Recently, \cite{gf15} have proposed that the heliocliff is a division between what they term the ``hot" heliosheath and the ``cold" heliosheath, with the heliopause lying at a greater distance.  The cold heliosheath is ``the region where escape [of higher energy particles (ACRs, pick up ions and the tail of the solar wind distribution)] across the heliopause occurs" (cf., energetic particle escape in the tail lobes) while the hot heliosheath contains compressed and heated solar wind, and may be the analog of the plasma sheet in the tail. A feature of the \cite{gf15} model is the decoupling of the solar wind plasma and ACRs, which is somewhat analogous to the free energetic particle streaming, unrelated to the local plasma flow, in the PSBL. 

\cite{s13} have interpreted the V1 observations in terms of a highly-structured heliopause including islands formed by the reconnection of heliospheric and interstellar fields.  Their Figure~4 is reminiscent of the structure of the plasma sheet in the geomagnetic tail, in particular at active times, showing the formation of neutral lines and plasmoids.  In their interpretation, the heliopause passes through the center line of these structures, and lies inside the region of fluctuating field and particle intensities (believed to be associated with the passage of magnetic islands) prior to the heliocliff, i.e., in the tail analogy, the heliopause lies in the plasma sheet rather than at the magnetopause.

As noted above, the unexpectedly high plasma densities following the heliocliff inferred from plasma wave observations \citep{gur13} have provided the main evidence that V1 may have entered the ISM.  In the case of the tail, lobe plasma (electron) densities measured by ISEE-3 were typically comparable to those in the plasma sheet \citep{z84}  ($\sim0.1$ cm$^{-3}$) and lower than those in the magnetosheath except near the magnetopause where magnetosheath plasma may enter the lobes.  The distribution of enhanced densities in the lobe also depends on the orientation of the interplanetary magnetic field  which controls reconnection at the magnetopause \citep{g85}.  Thus, by analogy, the ISM-like densities suggested by the V1 plasma wave observations might indicate that the spacecraft is in a region where interstellar plasma can enter the heliosphere along open field lines, and possibly is approaching the actual heliopause.  On this point, we note (as have, for example \cite{sm13} and \cite{gf15}) that a change in the direction of the ISM and/or the IMF at V1 (assuming it is still within the heliosphere) from the current configuration may have profound effects on the connection between the region beyond the heliocliff and the ISM, much as a change in the direction of the IMF impinging on the Earth, e.g., from southward (as in Figure~\ref{ms}), favoring dayside reconnection with magnetospheric fields, to northward, when reconnection of lobe field lines with the IMF may occur at the cusps (e.g.,  \cite{c83}), strongly influences the configuration of the magnetotail and its connection with the IMF.

\section{Acknowledgements}

The ISEE--3 EPAS (P.I., R.~J. Hynds, Imperial College) was designed and built by the Blackett Laboratory, Imperial College, London, The Space Research Laboratory, Utrecht, and the Space Science Department of ESA, ESTEC.  The ISEE-3 magnetic field data are from the Jet Propulsion Laboratory vector helium magnetometer (P.I. E.~J. Smith).  This paper was stimulated by the strong sense of {\it d\'ej\`a vu} generated by the recent Voyager~1 observations.

\section{Citations}
%\label{Section 3}

%Remember to give the title of all the articles in references and
%inclusive page numbers in each reference. This is a special
%requirement of ASR. We request authors to provide at least three 
%names before "et al." in multi-authored papers.

%\begin{itemize}
%\item Parenthetical: \verb|\citep{WB96}| produces \citep{WB96}.
%\item Textual: \verb|\citet{WB96}| produces \citet{WB96}.
%\item An affix and part of a reference:
   %\verb|\citep[e.g.][Ch. 2]{WB96}|
  % produces \citep[e.g.][Ch. 2]{WB96}.
%\end{itemize}

%%%%%%%%%%%%%%%%%%%%%%%%%%%%%%%%%%%%%%%%%%%%%%%%%%%%%%%%%%%%%%%%%%%%%%%%%%%%%
%% Appendices
% \appendix

\end{document}